\documentclass{aastex}
\usepackage{emulateapj5,epsfig}

\newenvironment{inlinefigure}{ 
\def\@captype{figure} 
\noindent\begin{minipage}{0.999\linewidth}\begin{center}} 
{\end{center}\end{minipage}\smallskip} 
\newcommand{\beq}{\begin{equation}}
\newcommand{\eeq}{\end{equation}}

\usepackage{graphicx}
\usepackage{amssymb}
\usepackage{epstopdf}
\font\tenbg=cmmib10 at 10pt

\def \rvecphi{{\hbox{\tenbg\char'036}}}

\begin{document}
\title{Stability of the Magnetopause of
Disk-Accreting Rotating Stars}

\author{
R.V.E. Lovelace\altaffilmark{1},
L. Turner\altaffilmark{2}, and
M.M. Romanova\altaffilmark{2}
 }

\altaffiltext{1}{Departments of Astronomy and  Applied and
Engineering Physics,
Cornell University, Ithaca, NY 14853-6801;
RVL1@cornell.edu}

\altaffiltext{2}{Department of Astronomy,
Cornell University, Ithaca, NY 14853-6801;
turner@cornell.edu; romanova@astro.cornell.edu}

\begin{abstract}

We discuss three  modes of oscillation of accretion disks 
around rotating magnetized neutron stars which may
explain the separations of the kilo-Hertz
-periodic oscillations (QPO) seen in
low mass X-ray binaries.   
  The existence of  these compressible, non-barotropic
magnetohydrodynamic (MHD)   modes requires 
that there be a maximum
in the angular velocity $\Omega_\phi(r)$ of the accreting material
larger than the angular velocity of the star $\Omega_*$,
and that the fluid is in
approximately circular motion near this
maximum rather than moving
rapidly towards the star or out of the disk plane
into funnel flows.   
    Our MHD simulations show
this type of flow and $\Omega_\phi(r)$ profile.   
   The first mode is a Rossby wave instability (RWI) mode which is
radially trapped in the vicinity of the maximum of 
a key function $g(r){\cal F}(r)$ at $r_{R}$.
    The real part of the angular  frequency of the mode 
is $\omega_r=m\Omega_\phi(r_{R})$, where 
$m=1,2...$ is the azimuthal mode number.
   The second mode,  is a mode driven by 
the rotating, non-axisymmetric component of the
star's  magnetic field.    It has an angular frequency
equal to the star's angular rotation rate $\Omega_*$.
   This mode is strongly excited near the radius
of the Lindblad resonance which is slightly
outside of $r_R$.
    The third mode arises naturally from the interaction of
flow perturbation with the rotating non-axisymmetric
component  of the star's magnetic field.  It has
an angular frequency $\Omega_*/2$. 
   We suggest that the first mode with $m=1$ is
associated with the upper QPO frequency, $\nu_u$;
  that the  nonlinear interaction of the first and
second modes gives the lower QPO frequency, $\nu_\ell =\nu_u-\nu_*$;
and  that the nonlinear interaction of the first and third modes gives
the lower QPO frequency
$\nu_\ell=\nu_u-\nu_*/2$, where $\nu_*=\Omega_*/2\pi$.

\end{abstract}


\section{Introduction}

 Low mass X-ray binaries often display twin
kilo-Hertz quasi-periodic oscillations (QPOs)
in their X-ray emissions (van der Klis 2006; Zhang
et al. 2006).
   A wide variety of different models have
been proposed to explain the origin and correlations
of the different QPOs.  
    These include
the beat frequency model (Miller, Lamb, \& Psaltis 1998;
Lamb \& Miller 2001; Lamb \& Miller 2003), the
relativistic precession model (Stella \& Vietri 1999),
the Alfv\'en wave model (Zhang 2004), and warped disk
models (Shirakawa \& Lai 2002; Kato 2004).

   A puzzling aspect of the some of the twin QPO
sources considered in this work is that
the  difference between the upper
$\nu_u$ and lower $\nu_\ell$ QPO frequencies is
roughly either the spin frequency of the star $\nu_*$
($3$ cases where $\nu_*=270,~330,~\&~ 363$ Hz) 
or one-half this frequency,
$\nu_*/2$ ($4$ cases where 
$\nu_*=401,~524,~581,~\&~619$ Hz), for the
cases where
$\nu_*$ is known, even though
$\nu_u$ and $\nu_\ell$ vary significantly
(see, e.g., Zhang et al. 2006). 
  A further type of behavior appears in
the source Cir X-1 (Boutloukos et al. 2006), but
this is not considered here.
  The cases where $\nu_u-\nu_\ell \approx \nu_*$
may be explained by the beat frequency
model (Miller et al. 1998), but the explanation
of the cases where $\nu_u-\nu_\ell \approx \nu_*/2$
is obscure.

       Li and Narayan (2004) analyzed
the stability of an incompressible rotating flow where
the radial profiles of the plasma angular rotation rate $\Omega_\phi$, 
the magnetic field $B_z$, and the
density $\rho$ change sharply at the magnetopause radius
but are independent of $z$.
The magnetic field was dynamically important in
the sense that
$\rho {\bf u}^2 \sim {\bf B}^2/4\pi$ with ${\bf u}=
r\Omega_\phi \hat{\rvecphi~}$.
    They found Rayleigh-Taylor and Kelvin-Helmholtz
type instabilities in the vicinity of the magnetopause.    
    Recently, Fu and Lai (2009) have done a systematic
analysis of short wavelength modes of a disk for cases where the
magnetic field is not dynamically important.  

   The present work is a continuation of the study of
Lovelace and Romanova (2007; hereafter  LR07) of the magnetohydrodynamic
(MHD) stability of the compressible, non-barotropic
stability of the boundary between an accretion disk
and the magnetosphere of a rotating magnetized star
for conditions where $\Omega_\phi$, 
 $B_z$, and
$\rho$ vary smoothly with radial distance but are
independent of $z$,
and where
the magnetic field is dynamically important.
    The radial profiles of these
quantities are known for different conditions from MHD simulations 
(Romanova et al. 2002;  Long, Romanova, \& Lovelace 2005;
Romanova, Kulkarni, \& Lovelace 2008; 
 Kulkarni \& Romanova 2008).
     The recent MHD studies (Romanova et al. 2008; Kulkarni
 \& Romanova 2008)   discovered conditions
where a global Rayleigh-Tayor instability occurs and changes
the nature of the accretion flow from a regular funnel flow
pattern to a chaotic flow of plasma fingers.   
    The present
work is concerned with a radially localized instablity of
a globally stable approximately axisymmetric flow.
   Of particular importance to the present work is 
that $\Omega_\phi(r)$ is observed to  go through a maximum
significantly larger than the angular rotation rate
of the star $\Omega_*=2\pi \nu_*$.
  The importance of the maximum of
$\Omega_\phi(r)$ for models of QPOs was discussed
earlier by Alpar and Psaltis (2005).
    LR07
showed that there was a Rossby wave instability (RWI) or
unstable corotation mode
with azimuthal mode number $m=1,2..$
radially trapped at $r_R$ near the maximum of $\Omega_\phi(r)$.
It was suggested that the upper QPO frequency was 
$\omega_r=m\Omega_\phi(r_R)$ and that
 the lower QPO frequency was due to the interaction of the
 mode with rotating non-axisymmetric field of the star.
    The  theory of the  RWI was developed by Lovelace et al. (1999) and  Li et al. (2000) for accretion disks and earlier by Lovelace \& Hohlfeld (1978) for disk galaxies. 
   Hydrodynamic simulations of the RWI
instability in disks were done by Li et al. (2001),  
while Sellwood \& Kahn (1991) used $N-$body simulations to study the
instability in galaxies.  
    The instability has important role in the accretion-ejection 
instability of disks discussed by Tagger and collaborators
(e.g., Tagger \& Varni\`ere 2006;  Tagger \& Pellat 1999).

  Section 2.1 develops the general compressible, non-barotropic 
 MHD  equations
for {\it  free} perturbations
an axisymmetric equilibrium flow with no $z-$dependence.
Section 2.2 treats the
{\it driven}
perturbations due to the rotating  non-axisymmetric component
of the star's magnetic field.  
   Section 2.3 treats the {\it driven-modulated} perturbations
which result from the interaction of the non-axisymmetric
field of the star {\it and} the flow perturbation.    
   Section  3 develops a detailed model of the
axisymmetric flow/field equilibrium based on results
from MHD simulations.
   Section 4 discusses the results obtained applying the
theory of \S 2 to the model of \S 3.   Section 4.1 
treats the free perturbations, \S 4.2 the driven
perturbations, and \S 4.3 the driven-modulated
perturbations.   Section 5 briefly discusses the
nonlinear effect of the unstable mode.
    Section 6 gives the conclusions of this work.

\section{Theory}

      We consider the stability of
 the magnetopause
of a rotating star with an aligned
dipole magnetic field.
   The envisioned geometry is shown
in Figure 1.   
   We use
an inertial cylindrical
$(r,\phi,z)$ coordinate system.
   The equilibrium  has
($\partial/\partial t=0$) and ($\partial/\partial \phi = 0$),
with the flow velocity
${\bf u} = u_\phi(r) \hat{\rvecphi~}=r\Omega_\phi(r) \hat{\rvecphi~}$.  
   That is, the accretion velocity $u_r$ and the vertical
velocity $u_z$  are assumed negligible compared with $u_\phi$.
      The equilibrium magnetic field
is  ${\bf B}=B(r)\hat{\bf z}$.
The equilibrium flow satisfies $\rho r (\Omega_K^2-\Omega_\phi^2)=
-d[p+B^2/(8\pi)]/dr$, where $\rho$ is the density, $p$ the pressure,
$\Phi$ the gravitational potential,
and $\Omega_K$ the Keplerian angular rotation rate of a single
particle.

\begin{inlinefigure}
\centerline{\epsfig{file=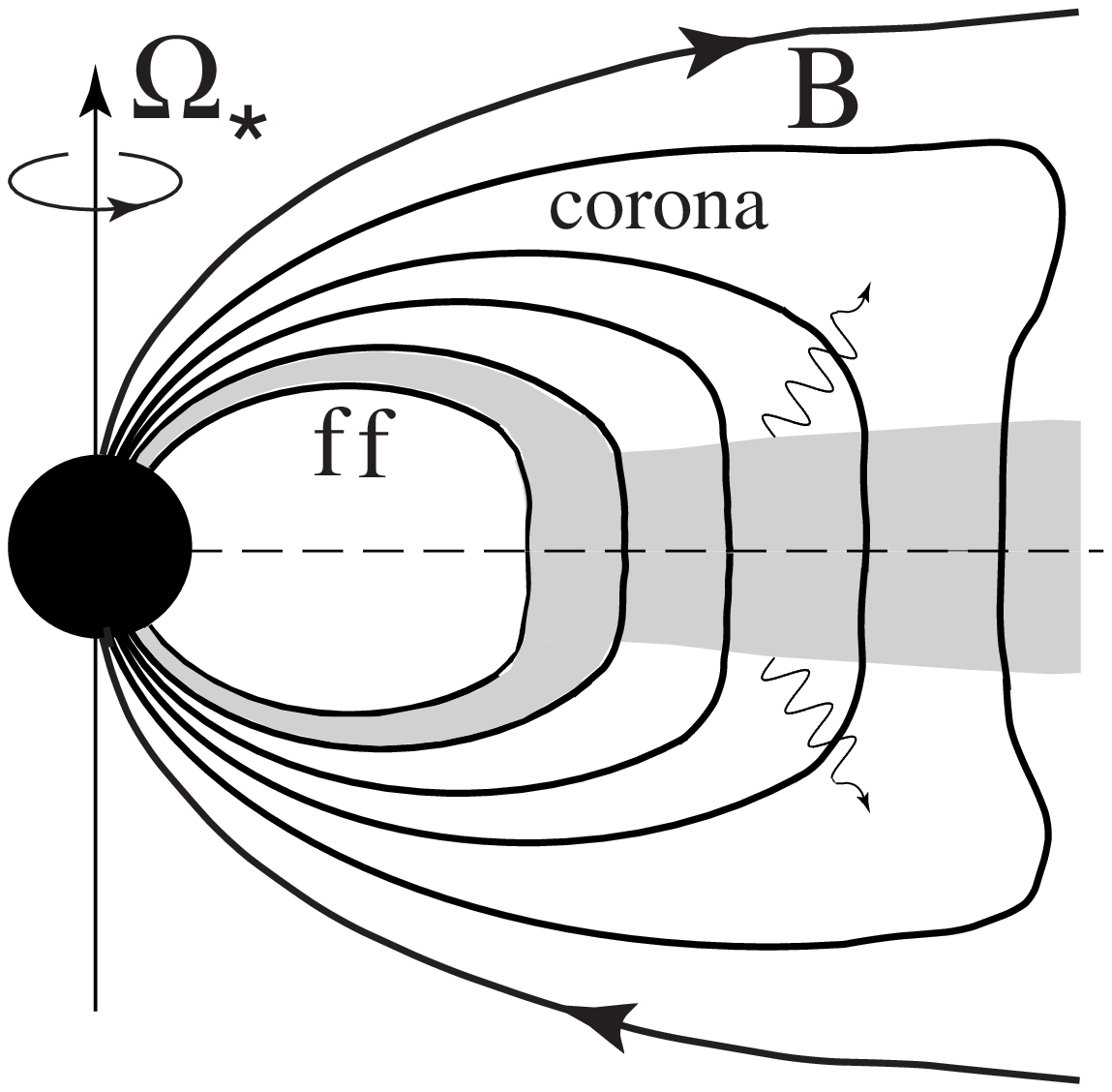,
height=2.5in,width=2.6in}}
\epsscale{0.8}
\figcaption{Sketch of the envisioned  disk,  magnetic
field, and rotating star geometry.  $\Omega_*$ is the angular rotation rate
of the star,  the wiggly lines indicates radiation from
the surface of the disk where the optical depth is unity, and 
ff indicates the funnel flow.   The region outside the disk and
funnel flow is occupied by a low-density, high-temperature corona.}
\end{inlinefigure}

\subsection{Free Perturbations}

The perturbed quantities are:
the density,
$\tilde{\rho} = \rho +
\delta \rho(r,\phi,t)$;
the pressure is
$\tilde{p} = p+\delta p(r,\phi,t)$;
and the  flow
velocity is $\tilde{\bf u} =
{\bf u} +\delta {\bf u}(r,\phi,t)$ with
${\bf \delta u} =
(\delta u_r,\delta u_\phi,0)$.
   Also, $\tilde{\bf B}=\hat{\bf z}B
+\hat{\bf z}\delta B$.
The equations for the
perturbed flow are
$$
{D \tilde{\rho}\over Dt}
+ \tilde{\rho}~ {\bf \nabla}\cdot
\tilde{\bf u} = 0~,
\eqno(1a)
$$
$$
{D \tilde{\bf u}\over Dt}  =
-{1\over \tilde{\rho}}
{\bf \nabla}\left( \tilde{p}+{B^2 \over 8\pi}\right)
 - {\bf \nabla}\Phi ~,
\eqno(1b)
$$
$$ 
{D S\over Dt} = 0~,
\eqno(1c)
$$
$$
{D \over Dt}\left({B \over \rho}\right) =0~,
\eqno(1d)
$$
$$
\nabla^2 \Phi = 4\pi  G \rho~,
$$
where $D/Dt \equiv \partial /\partial t
+ \tilde{\bf u}\cdot {\bf \nabla}$ and
we refer to $S \equiv {\tilde p}/(\tilde \rho)^\gamma$
as the entropy of the disk matter.

We consider perturbations $\sim f(r){\rm exp}
(im\phi - i \omega t)$, where $m=0,1,2,..$ is the azimuthal
mode number and $\omega$ the angular frequency.
For free perturbations $\omega=\omega_r+i\omega_i$ and for
the growing modes of interest $\omega_i >0$. 
(For the driven perturbations considered in \S 2.2 and
2.3,  $\omega_i=0$.)
 From equation (1a), we have
$$i\Delta \omega~ \delta \rho =
{\bf \nabla} \cdot
(\rho ~\delta {\bf u})~,
\eqno(2)
$$
where
$\Delta \omega(r) \equiv \omega - m \Omega_\phi(r)$ and
$\Omega_\phi = u_\phi/r$.

 From equation (1b) we have
$$ 
i\Delta \omega \delta u_r +
2 \Omega_\phi \delta u_\phi =
{1\over \rho}{\partial \delta p_* \over \partial r}
-{\delta \rho \over \rho^2} {d p_*
\over dr}
+{\partial \delta \Phi \over \partial r}~,
\eqno(3a)
$$
$$
i\Delta \omega \delta u_\phi -
{\Omega_r^2 \over 2 \Omega_\phi}
\delta u_r = ik_\phi
 {\delta p_* \over \rho}+ik_\phi \delta
\Phi~.~~~~~~~~~~~~~~
\eqno(3b)
$$
Here,
$\Omega_r \equiv
[r^{-3} d(r^4 \Omega_\phi^2)/dr]^{1\over 2}$
 is the radial
epicyclic frequency, $k_\phi \equiv m/r$
is the azimuthal wavenumber,
$$
p_* \equiv p+{B^2 \over 8\pi}~,\quad {\rm and}
\quad
\delta p_* \equiv \delta p 
+ {B\delta B \over 4 \pi}~.
$$
From equation (1c) and (1d), we have
$$
\delta p =c_s^2 \delta \rho 
-{i\rho c_s^2 \over \Delta \omega L_S}\delta u_r~,
~~ {\rm and}~~
\delta B ={\delta \rho \over \rho}B
-{i\rho\delta u_r \over \Delta \omega}
{d \over dr}\left({B\over \rho}\right)~.
$$
Combining these expressions,
$$
\delta p_*=(c_s^2+c_A^2)\delta \rho
-{i\rho \delta u_r \over \Delta \omega}
\left({c_s^2 \over L_S} + {c_A^2 \over L_B}\right)~,
\eqno(4)
$$
where 
$$ 
{1\over L_S} \equiv {1\over \gamma}{d \ln(S) \over dr}
~,\quad \quad
{1\over L_B} \equiv {d \ln(B/\rho) \over dr}~,
\eqno(5a)
$$ 
are the length-scales of the entropy $S\equiv p/\rho^\gamma$ and $B/\rho$
variations.
   Also, 
$$ 
c_s \equiv \left({\gamma p\over\rho}\right)^{1/2}~, \quad
{\rm and}
\quad c_A ={|B|\over\sqrt{4\pi \rho}}~,
\eqno(5b)
$$
are the sound and Alfv\'en speeds respectively.
  We denote $c_f\equiv (c_s^2+c_A^2)^{1/2}$ as
the fast magnetosonic speed.

    It is useful to introduce 
$$
\Psi \equiv {\delta p_*\over \rho}~.
\eqno(6)
$$
Equations (3) can then be written as
$$
A \delta u_r + B \delta u_\phi = f_r~,
\eqno(7a)
$$
$$
C \delta u_\phi + D \delta u_r = f_\phi~,
\eqno(7b)
$$
where 
$$
A = -i\left[\Delta \omega + {dp_*/dr \over
\Delta \omega  \rho L_*}
\right]~,
$$
$$
B= -2\Omega_\phi~,\quad C= -i\Delta \omega~,\quad 
D={\Omega_r^2 \over 2\Omega_\phi}~.
$$
Also,
$$
f_r=-{\partial \Psi \over \partial r}+{\Psi \over L_*}
-{\partial \delta \Phi \over \partial r}~,
$$
$$
f_\phi=-ik_\phi (\Psi +\delta \Phi)~,~~~~~~
$$
where
$$
{1\over L_*} \equiv {1\over \rho c_f^2}
{d p_* \over dr} - {1\over L_\rho} =
{1 \over c_f^2}\left({c_s^2 \over L_S}
+{c_A^2 \over L_B}\right)~,
\eqno(8a)
$$
and where
$$
{1\over L_\rho} \equiv {d\ln(\rho) \over dr}~.
\eqno(8b)
$$
For a strong magnetic field $c_A \gg c_s$
we have $L_* \rightarrow L_B$. 
For a weak magnetic field $c_A \ll c_s$,
we have $L_* \rightarrow L_S$.
  Using the equation for the equilibrium
$\rho r (\Omega_K^2-\Omega_\phi^2) = -dp_*/dr$
we have
$${1\over L_*}=-~{r(\Omega_K^2-\Omega_\phi^2) \over c_f^2}
-{1\over L_\rho}~,
\eqno(8c)
$$
which is useful later.

From here on we simplify the analysis by
neglecting the self-gravity of the disk.
That is, we neglect the $\delta \Phi$ terms 
in $f_r$ and $f_\phi$.

    We solve equations (7a) and (7b) to
obtain
$$
\rho \delta u_r =i{\cal F}
\left[{\Delta \omega \over \Omega_\phi}
\left({\partial \Psi \over \partial r} 
-{\Psi \over L_*}\right)-2 k_\phi \Psi \right]~,
\eqno(9a)
$$

$$
\rho \delta u_\phi ={\cal F}
\left\{ {\Omega_r^2 \over 2 \Omega_\phi^2}
\left({\partial \Psi \over \partial r} 
-{\Psi \over L_*}\right) \right.~~~~~~~~~~~~~~~
$$
$$
~~~~~~~~~~~\left.-k_\phi\left({\Delta \omega \over \Omega_\phi}
+{d p_*/d r 
\over \rho \Omega_\phi \Delta \omega L_*}
\right)\Psi \right\}~.
\eqno(9b)
$$
Here, 
$$
{\cal F} \equiv {\rho \Omega_\phi \over {\cal D}}~,
\eqno(10a)
$$
where
$$
{\cal D}=\Omega_r^2- (\Delta \omega)^2 -{dp_*/dr
\over \rho L_*}~,
\eqno(10b)
$$
is the `Lindblad factor.'
For real frequencies $\omega$, ${\cal D}(r)$ is
equal to zero at a Lindblad resonance at $r_L$.
On the other hand at a corotation
resonance $\Re[\Delta \omega(r)]=0$ at $r_R$.
  Using equation (8c) we find
$$
{\cal D}=\Omega_r^2- (\Delta \omega)^2 -
{r^2(\Omega_K^2-\Omega_\phi^2)^2 \over c_f^2}
-{r(\Omega_K^2-\Omega_\phi^2)\over L_\rho}~.
\eqno(10c)
$$  

Equations (9) can now be used to obtain
$$
\nabla \cdot (\rho \delta{\bf v})=
\overbrace{{i\Delta \omega \over r}
{\partial \over \partial r}\left({r{\cal F} \over \Omega_\phi}
{\partial \Psi \over \partial r}\right)}
+{i{\cal F} \over \Omega_\phi}
{\partial \Delta \omega \over \partial r}
{\partial \Psi \over \partial r}
$$
$$
-{i \over r} \left[{\partial \over \partial r}
\left({r {\cal F} \Delta \omega \over \Omega_\phi L_*}\right)
\right] \Psi
-{i{\cal F} \Delta \omega \over \Omega_\phi L_*}{\partial \Psi
\over \partial r}
-2ik_\phi {\partial {\cal F}\over \partial r} \Psi
$$
$$
-2ik_\phi{\cal F} {\partial \Psi \over \partial r}
-{ik_\phi {\cal F} \Omega_r^2 \over 2 \Omega_\phi^2 L_*} \Psi
+{ik_\phi{\cal F}\Omega_r^2 \over 2 \Omega_\phi^2}
{\partial \Psi \over \partial r}
$$
$$
-ik_\phi^2{\cal F} \left[ {\Delta \omega \over \Omega_\phi}
+{dp_*/dr \over \rho \Omega_\phi \Delta \omega L_*}
\right]
\Psi~.
\eqno(11)
$$
It is useful to note
that
$$
{\partial \Delta \omega \over \partial r}=
-~{k_\phi (\Omega_r^2-4\Omega_\phi^2) \over 2 \Omega_\phi}~.
$$

   From equations (4) and (9a) we
have
$$
\delta \rho= {\rho \Psi \over c_f^2}
-{{\cal F} \over \Delta \omega L_*}
\left[{\Delta \omega \over \Omega_\phi}
\left({\partial \Psi \over \partial r}-{\Psi \over
L_*}\right)-2k_\phi\Psi\right].
\eqno(12)
$$
Equations (11) and (12) can now be substituted into
equation (2).  
   When this is done we find that all of
the terms involving $\partial \Psi/\partial r$
apart from the overbraced term in equation (11)
cancel out.

Combining equations (9) with equation (2)
gives
$$
{1\over r}{\partial \over \partial r}
\left({r{\cal  F}\over \Omega_\phi}{\partial \Psi \over
\partial r}\right) =
$$
$$
\left[{\rho \over c_f^2}+  {k_\phi^2{\cal  F} \over \Omega_\phi}
+{1\over r}{\partial \over \partial r}
\left({r{\cal F} \over \Omega_\phi L_*}\right)
+{{\cal F} \over \Omega_\phi L_* ^2}
\right]\Psi 
$$
$$
+~\left[2k_\phi{\cal F}{d \ln( g{\cal F}) \over dr}\right]
{\Psi \over \Delta \omega}
+~\left[{k_\phi^2(dp_*/dr){\cal F}
\over \rho \Omega_\phi L_*}\right]{\Psi \over (\Delta \omega)^2}~,
\eqno(13)
$$
where $g=\exp(2\int dr/L_*)$.
In the limit of no magnetic field ($c_A \rightarrow 0$ and
$L_* \rightarrow L_S$), this equation becomes the
same as the equation for the Rossby wave instability
(Lovelace et al. 1999; Li et al. 2000).

A  quadratic form can be obtained
by multiplying equation (13) by $\Psi^*$
(the complex conjugate of $\Psi$)  and
integrating over the disk.
Assuming $r\Psi^*(d\Psi/dr)
{\cal F}/\Omega_\phi \rightarrow 0$
for $r\rightarrow 0,~\infty$, we obtain
$$
-\int d^2r~{{\cal F}\over \Omega_\phi}
\left(\left| {\partial \Psi \over \partial  r}\right|^2 + k_\phi^2 |\Psi|^2\right)
=\int d^2r~ {\rho \over c_f^2} |\Psi|^2
$$
$$
+\int d^2r~ \left[{1\over r}{\partial 
\over \partial r}\left( { r{\cal F} 
\over\Omega_\phi L_*}\right) +{{\cal F}\over
\Omega_\phi L_*^2}
\right] |\Psi|^2
$$
$$
+~2\int d^2r~k_\phi {\cal F}{ d\ln(g{\cal F}) \over dr}
 {|\Psi|^2 \over \Delta \omega}
$$
$$
+\int d^2r~ {k_\phi^2(dp_*/dr){\cal F}
\over \rho \Omega_\phi L_*}
 {|\Psi|^2\over (\Delta \omega)^2}~.
\eqno(14)
$$
For  corotation modes where 
$|\Delta \omega|^2 \ll \Omega_\phi^2$, 
the third and fourth integrals on the right-hand side
of equation (14) are possibly important.  
   Earlier work by Lovelace and Hohlfeld (1978),
Lovelace et al. (1999), and Li et al. (2000) 
found that a corotation instability
was  possible if the quantity $g{\cal F}$ has a
maximum or minimum at the radial distance $r_R$  where 
${\Re}(\Delta \omega) = 0$.  
      For a weak magnetic
field ($c_A^2 \ll c_s^2$ where $L_*\approx L_S$),
we find 
$g{\cal F} =S^{2/\gamma}{\cal F}$ which agrees
with the result of Lovelace et al. (1999) and
Li et al. (2000) where the self-gravity of the
disk was neglected.  
    With negligible self-gravity, 
instability is found only for conditions where
$g{\cal F}$ has a {\it maximum} as a function
of $r$.   For a maximum the radial group velocity
of the Rossby waves is directed towards $r_R$
(Lovelace et al. 1999).
    In the
strong $B-$field limit ($c_A^2 \gg c_s^2$ where
$L_*\approx L_B$), 
$g{\cal F} =(B/\rho)^2{\cal F}$.

We can rewrite equation (13) in a form
more amenable to numerical analysis.  That is,
$$
\Psi^{\prime \prime}
=\left({\overline{{\cal D}}^\prime \over \overline{{\cal D}}}
\right)\Psi^\prime
+\left(C_0+{C_1\over \Delta \omega}
+{C_2\over (\Delta \omega)^2}\right)\Psi~,
\eqno(15)
$$
where $\overline{{\cal D}}\equiv{\cal D}/(r\rho)
=(\Omega_\phi/r){\cal F}^{-1}$, where ${\cal D}$ is
given by equation (10b), and ${\cal F}$ by 
equation (10a).  The primes denote
radial derivatives. 
   Also,
$$
C_0 = k_\phi^2 +{{\cal D} \over c_f^2}-
{\overline{{\cal D}}^\prime \over L_* \overline{{\cal D}}}
+{1-L_*^\prime  \over L_* ^2}~,
$$
$$
C_1=2k_\phi\Omega_\phi\big[\ln(g{\cal F})\big]^\prime~,~~~~
C_2=k_\phi^2{p_*^\prime \over \rho L_*}~,\eqno(16)
$$
where $g$ is defined below equation (13).

If we let $\Psi = \big(\overline{{\cal D}}\big)^{1/2}\varphi $, then
equation (15) can be written in the form of a Schr\"odinger
equation,
$$
\varphi^{\prime \prime}=
\left(\overline{C}_0+{C_1\over \Delta \omega}
+{C_2\over (\Delta \omega)^2}\right)\varphi \equiv U(r)\varphi~,
$$
$$
\overline{C}_0 \equiv C_0+{3\over 4}\left({\overline{{\cal D}}^\prime \over \overline{{\cal D}}}\right)^2-{1\over 2}\left({\overline{{\cal D}}^{\prime\prime}\over \overline{{\cal D}}}\right)~,
\eqno(17)
$$
where $U(r)$ is an effective potential.  
 Both $U$ and $\varphi$
are in general complex for complex $\omega$.

    A quadratic form analogous to equation (14) 
can be obtained by multiplying equation (17) by
$\varphi^* ~dr$ and integrating over $r$.
Assuming $\varphi^*\varphi^\prime$ vanishes
at small and large $r$, we obtain
$$
-\int dr~ |\varphi^\prime |^2 = \int dr ~{\overline C}_0 |\varphi |^2
+2\int dr~k_\phi \Omega_\phi{[\ln(g{\cal F})]^\prime    \over \Delta \omega}| \varphi |^2
$$
$$
+\int dr ~{C_2\over (\Delta \omega)^2} |\varphi|^2~.
\eqno(18)
$$
The imaginary part of this equation gives
$$ 0 = \ldots  -\omega_i\int dr {\big[\ln(g{\cal F})\big]^\prime
\over (\Delta \omega_r)^2+\omega_i^2} |\varphi|^2 + \ldots~,
\eqno(19)
$$
where the ellipsis allow for contributions from the $\overline{C}_0$
and $C_2$ terms.
  If these terms are negligible as found in our numerical
evaluations,
it follows that a non-zero growth rate $\omega_i$ is possible
only for conditions where $[\ln(g{\cal F})]^\prime $ changes
sign as mention above (Lovelace \& Hohlfeld 1978).    
   Thus, $g{\cal F}$ is a key
function for stability of the considered flow.

    Using equations (4), (9a), and (12) we obtain
the relation of the temperature perturbation to $\Psi$,    
$$
{\delta T \over T} = A_0 \Psi + A_1 \Psi^\prime~,
$$
$$ 
A_0={\gamma-1 \over c_f^2} +{1 \over {\cal D} }
\left({\gamma-1 \over L_*^2}-{\gamma \over L_S^2}\right)
+{2k_\phi \Omega_\phi \over {\cal D}\Delta \omega}
\left({\gamma-1 \over L_*}-{\gamma \over L_S}\right),
$$
$$
A_1={1\over {\cal D}}\left({\gamma \over L_S}-{\gamma-1 \over L_*}\right)~.
\eqno(20)
$$
This equation shows that the temperature perturbation is
strongly enhanced at a corotation resonance where
$|\Delta \omega|$ becomes very small and at a
Lindblad resonance where ${\cal D}=0$.

\subsection{Driven Perturbations}

    The influence of a small non-axisymmetric component of the
stellar magnetic field can be studied by including in equation (1b)
the small force due to this non-axisymmetry.
        At a given radial distance, the total vertical magnetic field 
 $B=B^v +B^i$ consists of the vacuum component
 $B^v$ due to current flow inside the star and the
 induced field $B^i$ due to the current flow
 in the plasma outside the star.  
    The vacuum field
 field has the general form 
 $$
 B^v=B^{v0}(r)+\Delta B^v(r,\phi,t)~,
 $$
 $$
 \Delta B^v=B^{v1}(r)\exp(i\phi-i\Omega_*t) 
 +B^{v2}(r)\exp(2i\phi-i\Omega_*t)+...~,
 \eqno(21)
 $$
 where $B^{v0}$ is the axisymmetric component, 
 $B^{v1}\ll B^{v0}$ is the quadrupole component, and
 $B^{v2}\ll B^{v0}$ is the octupole component.  
   The quadrupole term can represent a non-centered but
aligned dipole field in the star.   
    The total magnetic force or acceleration is
${\bf F} = -{\bf \nabla} \big\{[\langle B\rangle
+\Delta B^v]^2\big\}/(8\pi \rho)$, (because
$({\bf B}\cdot \nabla ){\bf B}$=0), where $\langle B\rangle=
\langle B^i+ B^v \rangle$ and where the average is over $\phi$.
     Linearization
gives $\delta{\bf F}=
 -{\bf \nabla} [\langle B\rangle \Delta B^v(r,\phi,t)]/(4\pi \rho)$.
    Thus we have
$$
(\delta F_r,~\delta F_\phi)=
(\delta F_{r0},~\delta F_{\phi 0})\exp(im\phi -i\Omega_* t)~,
\eqno(22)
$$
where $m=1$ or $2$ and $\Omega_*$ is
the angular rotation rate of the star.   
    The calculation of \S 2.1 is modified slightly beginning
with equations (7) which are here replaced by
$$
A \delta v_r + B \delta v_\phi = f_r + \delta F_r~,
$$
$$
C \delta v_\phi + D \delta v_r = f_\phi+\delta F_\phi~,
\eqno(23)
$$
where $A,..,D$ are the same as defined below equation (7).

    The forcing term $\delta {\bf F}$ gives rise to
an additional contribution to $\rho \delta {\bf v}$
not included in \S 2.1.   This contribution is
$$
\rho \delta v_r ={1 \over r\overline{\cal D}}( -C \delta F_r+B \delta F_\phi) \equiv H_r~,
$$
$$
\rho \delta v_\phi={1 \over r\overline{\cal D}}(D \delta F_r- A \delta F_\phi) \equiv H_\phi~,
\eqno(24)
$$
where $\overline{\cal{D}}={\cal D}/(r\rho)$ and ${\cal D}$ is the Lindblad factor defined in equation (10b).
   Including this contribution we find that response of the
flow $\varphi =\Psi/ \big(\overline {\cal D}\big)^{1/2}   $ is
given by
$$
\varphi^{\prime \prime} -U(r) \varphi =
i{r \big(\overline {\cal D})^{1/2} \over \Delta \omega}
{\bf \nabla}\cdot {\bf H}~,
\eqno(25)
$$
where $U$ is defined in equation (17) and ${\bf H}$
is a known function determined by $\Delta B^v$.
  We are interested in only the inhomogeneous solutions
to this equation where  $U, ~\omega,~\Delta \omega,$ and
$\overline{\cal{D}}$ are real.

\subsection{Driven-Modulated Perturbations}

      Here we consider  the case where 
the abovementioned  driving force $\delta {\bf F}$
is modulated by the flow perturbation $\Psi$
(LR07). 
   This modulation comes about naturally
by including the magnetic field perturbation
$\delta B$ in the magnetic force ${\bf F}$.
    That is, 
${\bf F} = -{\bf \nabla} \big\{[\langle B\rangle+\delta B
+\Delta B^v]^2\big\}/(8\pi \rho)$.
     Linearization  of this force gives a contribution
${\bf F}^m = -{\bf \nabla} \big[\delta B\Delta B^v\big]
/(4\pi \rho)$.   
     The equations of \S 2.1 can be used to derive
an exact formula relating $\delta B$ to $\Psi$, but 
 here we use a simplified formula for one of
 the dominant terms,
$\delta B =B \Psi({\cal D}L_*^2)^{-1}$, in order
to limit the complexity of the equations.
     An equation for the  driven modulated perturbations   
is then obtained      
by the replacement in
equation (23) of $\delta {\bf F} \rightarrow
\delta {\bf F}^m =  \Psi^* \delta {\bf F}/({\cal D}L_*^2)$,
where $\delta {\bf F}$ is still given by equation (22)
and is a known function determined by $\Delta B^v$.
  The use of $\Psi^*$ rather than $\Psi$  is required 
due to our assumed dependence of $\delta {\bf F}$ in equation (22).  
   For the driven-modulated perturbations we obtain
in place of equation (25)
$$
\varphi^{\prime \prime} -\widehat{U}(r) \varphi =
i{r \big(\widehat{\overline {\cal D}}\big)^{1/2} \over \widehat{\Delta \omega} }
{\bf \nabla}\cdot\bigg\{\widehat{\bf M}\cdot
\bigg[\bigg({\widehat{\overline{\cal D}}^{-1/2}\varphi \over
r\rho L_*^2   }\bigg)^*
\delta {\bf F}\bigg]\bigg\}~,
\eqno(26)
$$
where we have used the fact that $\Psi=(\overline{\cal D})^{1/2}\varphi$.
The hats over different quantities indicates that they are 
now operators with $\omega \rightarrow i(\partial/\partial t)$
and $k_\phi \rightarrow -i(\partial/\partial \phi)$.
The matrix  $\bf{M}$ allows us to write equation (24)
as ${\bf H}={\bf M}\cdot \delta {\bf F}$.  

        Equation (26) is a
 linear equation for $\varphi$ so that   
 in general
$$
\varphi=\varphi_0\exp(-i\omega t) +\varphi_1\exp(i\phi-i\omega t)
+\varphi_2\exp(2i\phi-i\omega t)+..~,
\eqno(27)
$$ 
where $\omega$ is real but undetermined at this point.
     Clearly, the time and angle dependence on both sides of equation (26)
 must match.  

   For the case where there is a small quadrupole field component
$\Delta B =B^{v1}\exp(i\phi-i\omega t)$,
equation (26) implies
$$
\varphi_0^{\prime \prime} -U_{0,\omega}(r) \varphi_0 =\big[{\rm Op}(\varphi_1^*) \big]_{0,\omega}~~~~ ~(m=0),\quad
$$
$$
\varphi_1^{\prime \prime} -U_{1,\omega}(r) \varphi_1 =\big[{\rm Op}(\varphi_0^*) \big]_{1,\omega}~\quad (m=1),
\eqno(28)
$$
where ${\rm Op}$ stands for the linear operator 
on $\varphi^*$ on   the right-hand
side of equation (26) and the new subscripts indicate the
$m$ value and the $\omega$ value.   Here, we necessarily
have 
$$
\omega = {\Omega_*\over 2}~,
\eqno(29)
$$
 as found earlier by LR07.

   For the case where there is a small octupole field component
$\Delta B=B^{v2}\exp(2i\phi-i\omega t)$,
equation (26) implies
$$
\varphi_1^{\prime \prime} -U_{1,\omega}(r) \varphi_1 =\big[{\rm Op}(\varphi_1^*) \big]_{1,\omega}~,
\eqno(30)
$$
where we again have $\omega=\Omega_*/2$.

   In contrast with the case of driven perturbations of \S 2.2,
equation (26) is linear in $\varphi$ so that its magnitude is
indeterminate.   
   Further study is needed to determine the
magnitude of $\varphi$ in this case.   One possibility is
to generalize equation (26) to include on the right-hand-side
of the equation the inhomogeneous
 driving force $\propto \delta {\bf F}$ 
as well the nonlinear force  $\delta {\bf F}^{nl} = -{\nabla}
\big(|\delta B |^2\big)/(8\pi \rho)$.
     Owing to the nonlinearity
the driving force $\delta {\bf F}$ at frequency $\Omega_*$ 
can  generate the $1/2$ subharmonic 
oscillations of $\varphi$ at the frequency 
$\Omega_*/2$ described by equation (26).   
  Analogous subharmonic generation is known in similar
systems (e.g., Minorsky 1974).

\section{Model of Equilibrium}

     Figure 2 shows sample measured profiles of the midplane axial
magnetic field ($B$),  the azimuthal frequency of the
disk matter [$\nu_\phi = v_\phi/(2\pi r)$] and 
its density ($\rho$) from MHD simulations by Romanova,
Kulkarni, \& Lovelace (2008) for globally stable case.
 The simulations were
three dimensional for an approximately axisymmetric case.  
These profiles motivate our choice of  analytic functions
to represent these quantities.

   The fact that $\nu_\phi(r)$ decreases as $r$ decreases
close to the star is due to  magnetic braking.   
  A small
twist of the star's magnetic field  transports angular momentum
of the disk matter to the star.   
   That is, for  $z>0$ there is a 
vertical flux of angular momentum
$-rB_\phi B_z/(4\pi) >0$ with $| B_\phi | \ll |B_z|$ which 
transports the disk angular momentum to the star along
the star's field lines (see Figure 1).  
   This loss of angular momentum
implies a mass accretion rate of the disk (outside
of the region of the funnel flow) of 
$\dot{M}_d = -r^2(B_\phi)_h B_z/(d \ell /dr)$ ($=$ const)  in
the absence of viscosity (see Lovelace, Romanova, \& Newman 1994), 
where the $h-$subscript
indicates evaluation at the top surface of the disk, $\ell =ru_\phi$
is the specific angular momentum, and $d \ell/dr$
is positive for the considered profiles.  
  The turbulent viscosity due to the magnetorotational 
instability (MRI) is absent in the region of the disk  where
$c_A>c_s$ and/or $d\Omega_\phi/dr >0$ (Balbus \& Hawley 1998).
Note however that the disk equilibria discussed below
neglect the accretion and have $B_\phi =0$.

\begin{inlinefigure}
\centerline{\epsfig{file=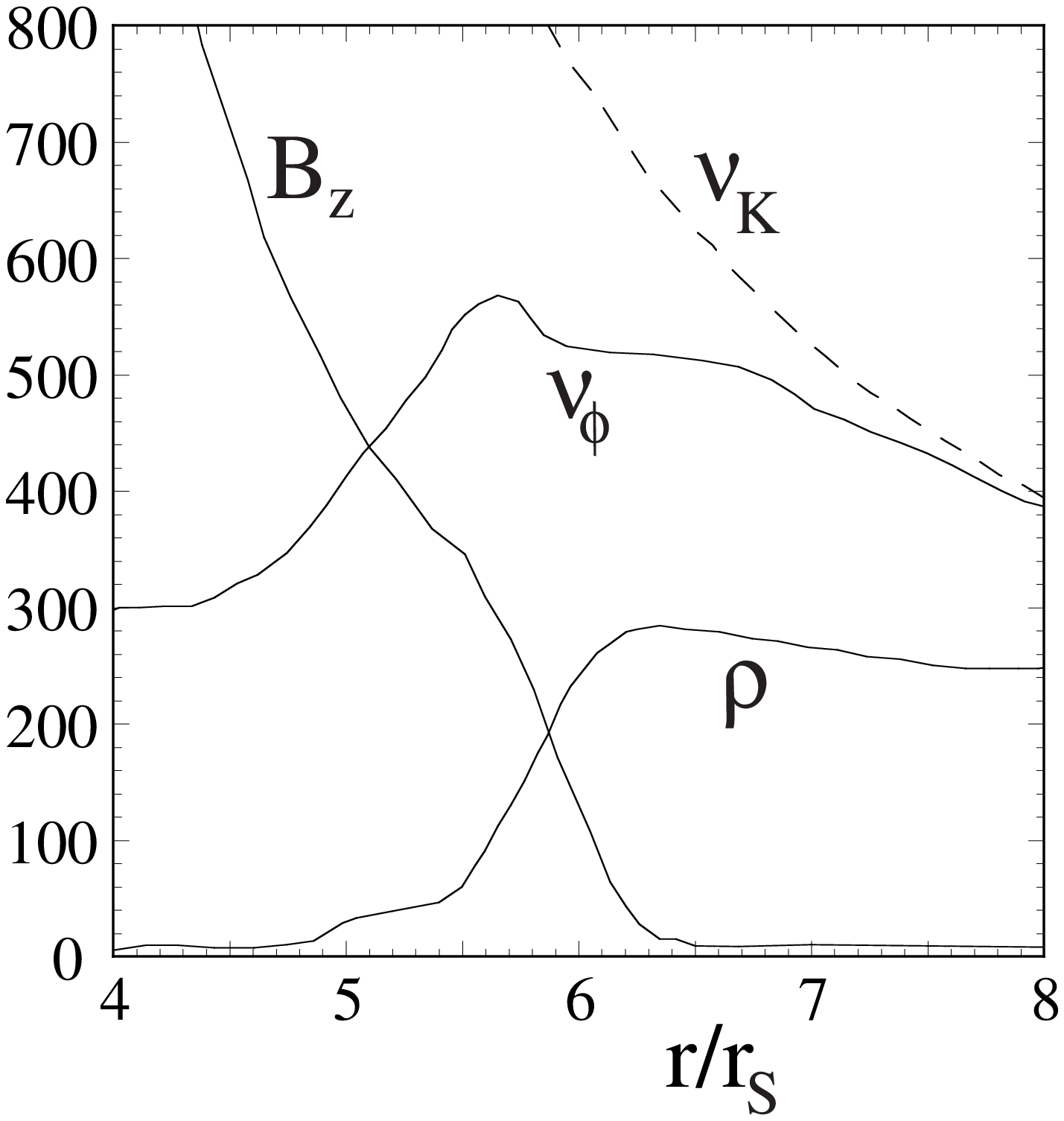,
height=2.9in,width=3.in}}
\epsscale{0.8}
\figcaption{Midplane radial profiles of the magnetic
field, the azimuthal frequency  [$\nu_\phi=v_\phi/(2\pi r)$],
the Keplerian frequency ($\nu_K$), and the
density $\rho$ from the MHD simulations. 
 The vertical scale is
for  the  azimuthal frequency of the disk matter;  the star's frequency is
$\nu_*=300$ Hz.  The scalings of $B_z$ and $\rho$ is arbitrary.}
\end{inlinefigure}

    We assume a  pseudo-Newtonian  
gravitational potential $\Phi_g=-GM_*/(r-r_S)$,
where $M_*$ is the star's mass and $r_S\equiv 2GM_*/c^2 =
4.14\times 10^5$ cm for a $1.4M_\odot$ star.
The angular
velocity of a  single paraticle  is $\Omega_K=2\pi \nu_K=
\{GM_*/[r(r-r_S)^2]\}^{1/2}$ for $r\geq 3r_S$. 
   Near the star,  the 
rotation frequency of the matter
is modeled following LR07 as
$$
\nu_\phi^0(r) ={\nu_* f(r) \over 1+f(r)}
+ {\nu_K(r) \over 1+f(r)}~,
\eqno(31)
$$
which is a first approximation as explained below.
Here,  $\nu_*$ is the rotation frequency of the star and
$f(r) \equiv \exp[-(r-r_0)/\Delta]$ with $r_0$
the standoff distance of the boundary layer
and $\Delta$ is its thickness.    The azimuthal
velocity of the matter is $v_\phi =2\pi r \nu_\phi^0$.
  Both $r_0$
and $\Delta$  are expected
to depend on the accretion rate and  the
star's magnetic field.

\begin{inlinefigure}
\centerline{\epsfig{file=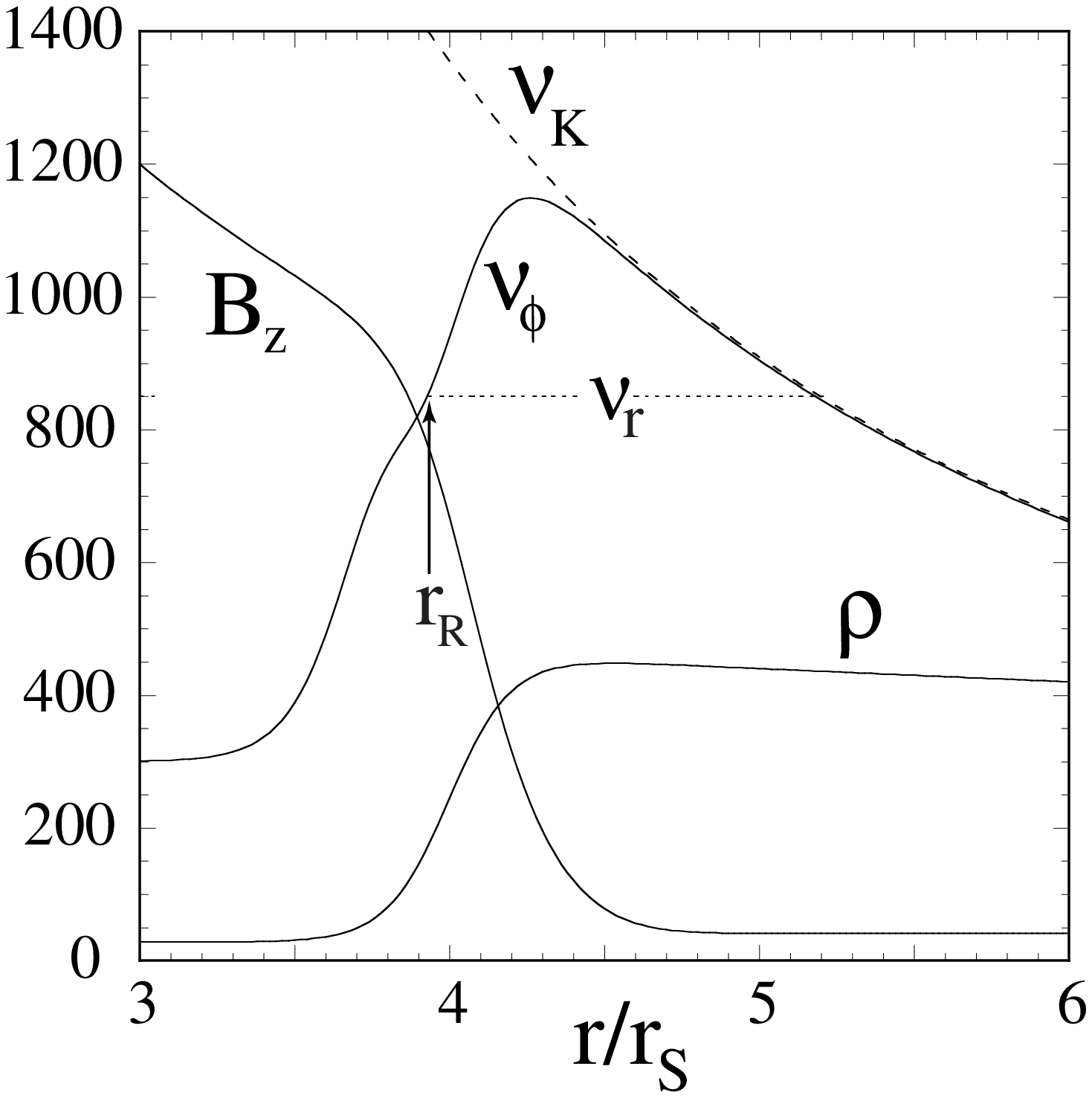,
height=2.9in,width=3.in}}
\epsscale{0.8}
\figcaption{Radial profiles of the magnetic
field ($B_z$), the azimuthal frequency  [$\nu_\phi=v_\phi/(2\pi r)]$,
the Keplerian frequency ($\nu_K$), and the
density $\rho$ of the model discussed in \S 3.   
The horizontal dotted line indicates the real part of
the frequency $\nu_r$ of the unstable $m=1$ corotation mode
discussed in \S 4.1.
 The vertical scale is
for  the  azimuthal frequency of the disk matter;  the star's frequency
is taken to be $\nu_*=300$ Hz.    
    The plot scales for  $B_z$ and $\rho$ are arbitrary.
For this case, $r_0/r_S =4$ and $\Delta/r_S =0.1$ in equation
(25), $\epsilon=0.05$ and
$\rho_0=0.013$ g cm$^{-3}$ in equation (26), $\gamma=5/3$, and
at $r/r_S=6$, $c_A/c_s = 0.156$. 
For this plot $B_z(r=r_0)=9.32\times 10^8$ G. }
\end{inlinefigure}

  The radial force equilibrium for the  axisymmetric flow is 
$$
\rho r(2\pi)^2(\nu_K^2-\nu_\phi^2) 
= -~{d\over dr}\left(p+{B^2 \over 8\pi}\right)~,
\eqno(32)
$$
where $p=\rho c_s^2/\gamma$ is the pressure in
the disk and $c_s$ is the sound speed. 
  The density profile is modeled as
$$
\rho=\rho_0 \left(\epsilon+{(1-\epsilon)\exp(-0.05 r/r_S) \over 1+f(r)}\right)~,
\eqno(33)
$$
with $f(r)$ the same function as in equation (31) and $\epsilon$ is a
positive quantity much less than unity.
   The sound speed is modeled by choosing say $c_{sK} =0.1 r \Omega_K(r)$
and then letting
$$
c_s ={c_{sK}(r_0) f(r) \over 1+f(r)}+{c_{sK}(r) \over 1+f(r)}~.
\eqno(34)
$$
At large distances $r\gg r_0$, $c_s/v_\phi =0.1$  corresponds
to a disk half-thickness $h \approx 0.1r$.  Inside of $r_0$ the
sound speed is assumed to be a constant.
Using these expressions we    first solve equation (32) for $B^2$
using equation (31) and neglecting the pressure gradient.  
We then go back and obtain the correction $\delta \nu_\phi$ to the azimuthal
frequency needed to account for the pressure gradient;
that is, $2\rho r(2\pi)^2 \nu_\phi^0 \delta \nu_\phi =dp/dr$ so that the
actual rotation frequency is $\nu_\phi =\nu_\phi^0 + \delta \nu_\phi$.
The correction is small, $|\delta \nu_\phi/\nu_\phi^0| \sim (c_s/v_\phi)^2
\ll 1$.  The radial epicyclic frequency is calculated from $\Omega_r=2\pi \nu_r$ with $\nu_r^2=r^{-3}d(r^4 \nu_\phi^2)/dr$.

     Representative curves are shown in Figures 3 and 4.    
 The value of the magnetic field at $r/r_S =6$ is arbitrary, but
 here it is  chosen so that $c_A < c_s$ which allows the
 MRI  to grow  which in turn gives rise
 to a turbulent viscosity in the disk (Balbus \& Hawley 1998).
 The maximum of
 the disk frequency $\nu_\phi$ is about $1149(4r_S/r_0)^{1.68}$ Hz
 for $\Delta/r_S =0.1$, and it occurs at a distance about $0.26 r_S$  larger
 than $r_0$.   For $\Delta/r_S =0.2$, the maximum of $\nu_\phi$
 is somewhat smaller and it occurs at a distance about $0.4r_S$
 larger than  $r_0$.

\begin{inlinefigure}
\centerline{\epsfig{file=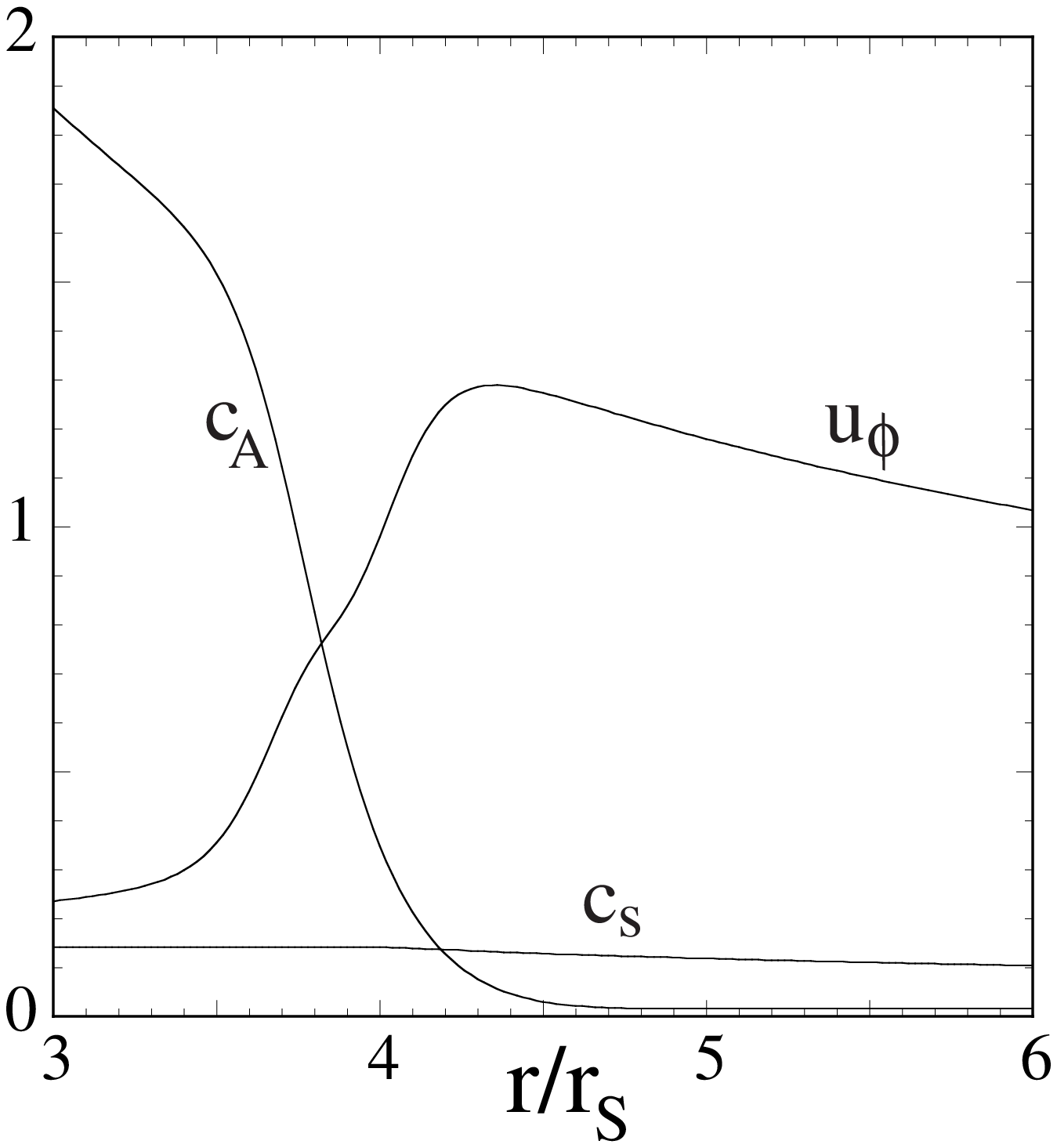,
height=2.9in,width=3.in}}
\epsscale{0.8}
\figcaption{Radial dependences of the Alfv\'en speed,
$c_A$, sound speed $c_s$, and azimuthal
velocity $u_\phi$ for  the same
conditions as in Figure 3.   In this plot
the speeds are in units of $10^{10}$ cm/s. }
\end{inlinefigure}

 \section{Results}
 
     We solve the equations of \S 2 using the equilibrium
profiles of \S3.  For this we use Maple V. 12 where we
define about $40$ different functions of $R=r/r_S$,
for example, $\rho(R), \nu_\phi(R), \nu_r(R), 
\Delta\omega(R), B(R), .. ,U(R)$.  
Some of the functions, for example, $U$, are complex
for complex $\omega$.    The axisymmetric
($m=0$) perturbation are found to be {\it stable} if the width
of the boundary layer is not too narrow; that is,
there is stability for $\Delta/r_S \gtrsim 0.02$.
In the following we consider the possible instability
of modes with $m=1,2,...$.
 
 \subsection{Free  Perturbations}
 
     Figure 5 shows the radial dependence of
the real part of the key function $g(r){\cal F}(r)$   for the same
conditions   as Figure 3.  This function has a
maximum at $r_R=3.92r_S$ indicated by the
vertical arrow marked corotation resonance.
The function changes sign at $r_L=4.07r_S$ which
is a Lindblad resonance where ${\cal F}$ (and
${\cal D}$) change signs.    
    The maximum of 
$\Re[g(r){\cal F}(r)]$ appears to be a general feature
for profiles similar to those in Figure 2.   
    The dependence
can be traced to the dependence of ${\cal D}(r)$ 
(equation 10b).    For $r$ decreasing significantly below 
$r_0$, the term $-(dp_*/dr)/(\rho L_*)$ in ${\cal D}$
 becomes increasingly negative.  This is due
 to the radial dependence of the magnetic pressure 
 and the small density.   Note $L_*$ is negative in
 this region.
On the other hand for $r$ increasing from $\sim r_0$,
the positive contribution of $\Omega_r^2$ begins
to dominate.  The combination of these dependences
gives a $\Re[g(r){\cal F}(r)]$ profile with a maximum
at a distance inside the Lindblad resonance as
shown in Figure 5.

\begin{inlinefigure}
\centerline{\epsfig{file=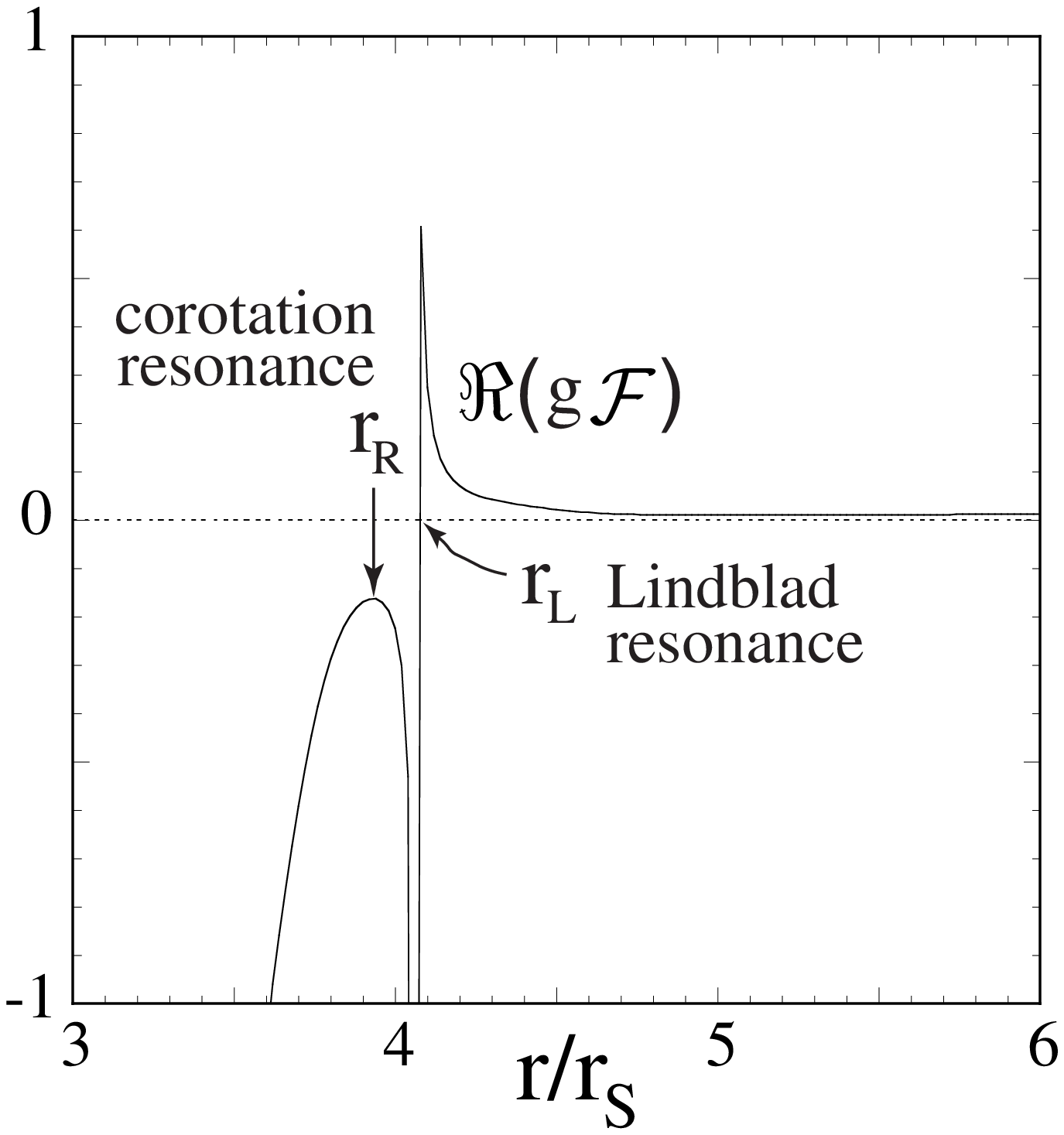,
height=2.9in,width=3.in}}
\epsscale{0.8}
\figcaption{Radial dependence of the real
part of the key function $g{\cal F}$ for
for the same conditions as for Figure 3.  
  The imaginary part of the function is
small compared with the real part.  
  The function has only a weak
dependence on $\omega$ which
for this plot is $\omega \approx 850+i89$ Hz.  
   We find maximum 
instability for 
$\Re[\Delta \omega(r)] = 0$ at the radial
location of the maximum of $\Re(g{\cal F})$
indicated by the vertical arrow.  The other
arrow shows a Lindblad resonance where
$\Re({\cal F})$ changes sign.
The vertical scale is arbitrary.
}
\end{inlinefigure}

       Figure 6 shows the real part of the
effective potential $U(r)$ for the same case
as Figure 5.   
  The real part of the frequency
is chosen to give $\Re[\Delta \omega(r)]=0$
at $r_R/r_S=3.92$ of the maximum of   $\Re[g(r){\cal F}(r)]$
because this give the maximum growth rate.   
   For $m=1$
this gives $\nu_r=\omega_r/2\pi =850$ Hz. 
   Clearly this frequency must be larger than 
rotation frequency of the star $\nu_*$.  
 
   The depth of the potential well shown in Figure 6
increases as the imaginary part
of the frequency $\omega_i=\Im(\omega)>0$  (the growth rate) decreases.  
    We use a WKBJ treatment with 
$\varphi \propto k^{-1/2} 
\exp\big(\pm i\int^r dr k\big)$ and $k=[-\Re(U)]^{1/2}$.   
  The imaginary part of $U$ is small compared with the real part. 
  The allowed values of $\omega_i$ are then
calculated using   the Bohr-Sommerfeld quantization 
condition, 
$$
\int_{r_{\rm in}}^{r_{\rm out}} dr~ k =
\left(n+{1\over 2}\right)\pi~,~~~~~ n=0,1,2,..~,
\eqno(35)
$$ 
where $k\equiv \sqrt{-\Re(U)}$, and  $r_{\rm in},~\&~r_{\rm out}$
are the radii where $\Re(U)=0$.  

     For the case
of Figure 6, $r_{\rm in}/r_S =3.83$ and
 $r_{\rm out}/r_S =3.97$.
The largest growth rate $\omega_i$  corresponds
to $n=0$ and for the case of Figure 6 with $m=1$ this gives
$\nu_i=\omega_i/2\pi = 89$ Hz so that $\omega_i/\omega_r \approx 10\%$.
For modes with $m\geq 2$  also grow with similar $\omega_i/\omega_r$
values but as discussed below these modes do not give
time variations in  the total flux from the source.

  For a more gradual boundary layer with $\Delta =0.2$ and $m=1$,
we find $\Re[\Delta \omega(r)]=0$ for $\nu_r=750$ Hz
at the radius  $r/r_S=3.83$ of the maximum of   $\Re[g(r){\cal F}(r)]$.
 From equation (35) we find $\nu_i =120$ Hz.
    For $\Delta$ increasing from $0.2$, we find that $\nu_r$ and $r_R$
continue to decrease gradually and $\nu_i$ also decreases.

\begin{inlinefigure}
\centerline{\epsfig{file=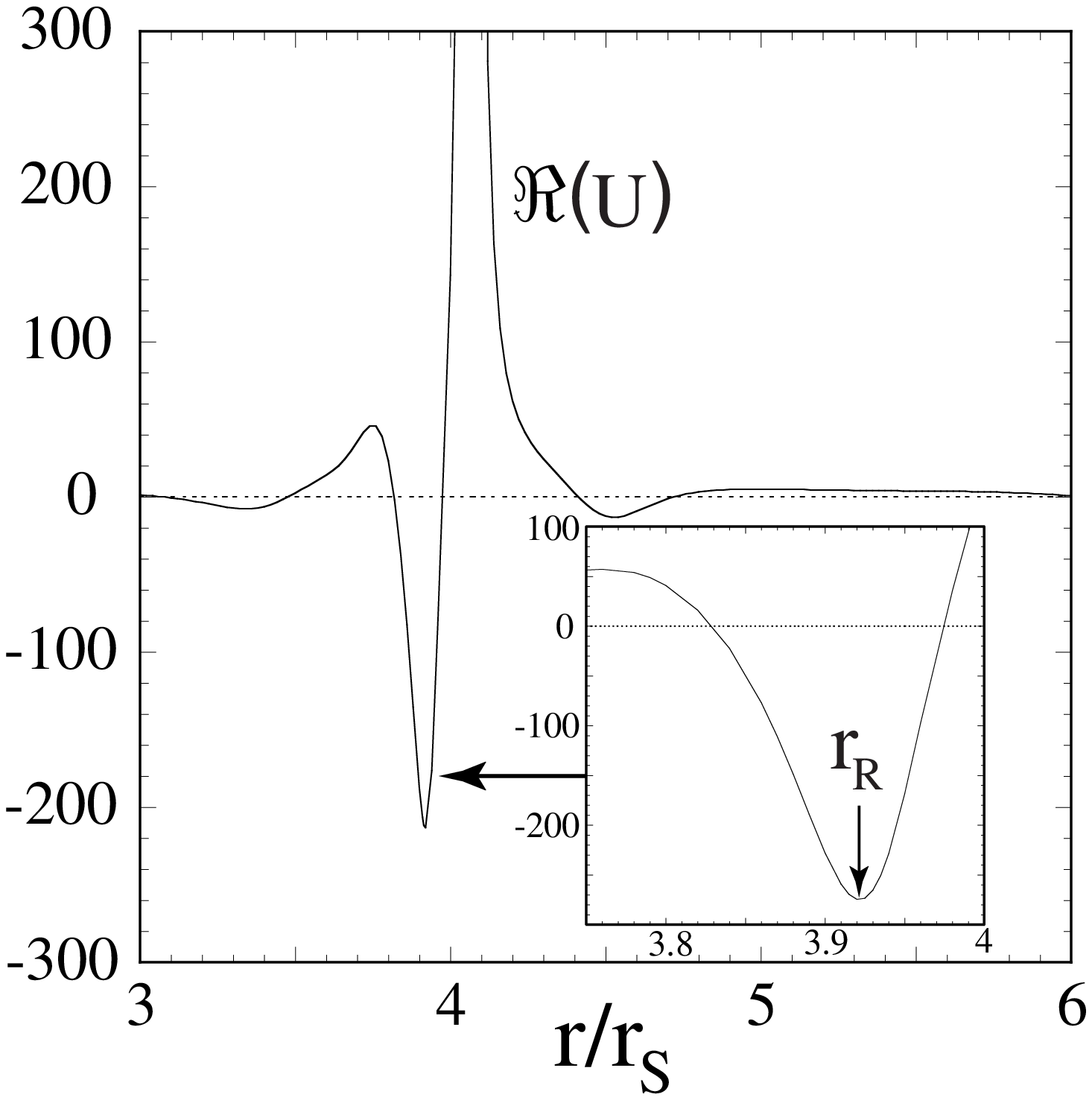,
height=2.9in,width=3.in}}
\epsscale{0.8}
\figcaption{Radial dependence of the real part
of the effective potential $\Re[U(r)]$ (multiplied by
$r_S^2$) for the same
conditions as for Figure 5. 
The imaginary part of $U$ is much smaller than 
the real part.
 }
\end{inlinefigure}

    Owing to the perturbation, the surface temperature of
the disk is 
$$
T(r,\phi,t)=
T_0+\Re\big[\delta T_1\exp(i\phi-i\omega_1 t)
$$
$$
+~\delta T_2(r)\exp(2i\phi-i\omega_2 t)+~.~.\big]~,
\eqno(36)
$$
where $T_0(r)$ is the unperturbed temperature,
$\delta T_{1,2}(r)\ll T_0$ are the amplitudes of
the $m=1,~2$ corotation modes, and $\omega_{1,2}$
are their frequencies.    
    We neglect the difference  between the midplane and the
surface temperature of the disk.    
The corresponding flux density from the disk
surface is proportional to
$S(r,\phi,t)\sim T_0^4+\Re\big[4T_0^3\delta T_1\exp(i\phi-i\omega_1t)
+4T_0^3\delta T_2\exp(2i\phi-i\omega_2 t)+..\big]$.
The total  flux for a face-on disk, $L\sim \int rdrd\phi S$,
is independent of time:
 the $\phi-$integration annihilates
the terms dependent on $\phi$. 
    For a more general disk
orientation, with the disk angular momentum
tilted say towards the line
of sight by an angle $\iota$, the Doppler effect due to the
disk rotation gives a boost to
the frequency for say $\phi=0$  and
a decrement for $\phi=\pi$.  
   This corresponds to multiplying $S$ by the
Doppler factor   
$D(r,\phi) \equiv [1+ \varepsilon(r) \cos(\phi)]^4
\approx 1 +4 \varepsilon \cos(\phi)$,  
where  $\varepsilon =[v_\phi(r)/c]\sin(\iota)$ and
$\varepsilon^2 \ll 1$.
   For the case of Figure 6
where $r=3.92r_s$, $v_\phi/c =0.289$.
    Consequently, there is
a contribution to the source
flux $\delta  L \sim \int rdr d\phi
D(r,\phi)S(r,\phi,t) \sim  
16\pi\Re\big[ \int rdr T_0^3\varepsilon \delta
T_1(r)\exp(-i\omega_1t)\big]$.  
     We interpret this frequency  as the
upper frequency component of the twin QPOs
as argued earlier  by LR07.
   Note that the higher order terms $\delta T_2,~\delta T_3,...$
give no contribution to the total flux due
to the $\phi-$integration.

\subsection{Driven Perturbations}

     Here we consider the  driven perturbations
of the flow which result from a quadrupole
or octupole component of the star's
magnetic field as discussed in \S 2.2.
   Inspection of equation (25) reveals that the radial
locations of the    Lindblad
resonances where ${\cal D}$ (or $\overline{\cal D}$) vanishes
are very important for excitation of the flow (LR07).
At such a resonance the right-hand-side equation (25) -
the driving term - is proportional to 
$\overline{\cal D}^\prime(\overline{\cal D})^{-3/2}$.

   We first consider the quadrupole field case $B^{v1}$  where $m=1$
for  $\nu_r=\omega_r/2\pi = 300$ Hz and $\omega_i =0$.
    We find that  $\overline{\cal D}$ goes 
through zero at one  radius, $r_L=4.07r_S$ 
for this case.
        Near this radius we find
  $\overline{\cal D}\approx a(x-bx^2)$, 
 where $x\equiv (r-r_L)/r_L$ and
 $a,~b >0$ are constants.   We are 
 interested as explained below to determine the
 amplitude of this driven mode near the vicinity
 of the corotation mode at $r_R=3.92r_S$.
     For this purpose we develop an approximate
 solution to equation (25) for $x^2 \ll 1$ by
retaining only the most
 singular term in the effective potential which is  
 $U\approx (3/4)\big(\overline{\cal D}^\prime/\overline{\cal D}\big)^2$.
 The right-hand side of the equation can be approximated
 as  $K{\overline{\cal D}}^\prime (\overline{\cal D})^{-3/2}$,
 where $K\propto B^{v1}$. 
   Thus equation (25) simplifies to
$$
   {d^2 \varphi \over dx^2} -\kappa^2\varphi=   
  { K (1-2bx)\over (x-bx^2)^{3/2}}~,
  \eqno(37)
$$
where 
 $\kappa^2  = (3/4)[ (1-2bx)/( x-bx^2)]^2$.
   The inhomogeneous solution to this equation for $x^2 \ll1$ is
$$
\varphi ={K\over 3  \sqrt{x}}~.
\eqno(38)
$$
Although $\varphi$ diverges at $x=0$, note
that $\Psi =\delta p_*/\rho$ is a constant.
   For $-x$ increasing we find that  the dependence
of    $\kappa^2$ changes from $\sim (3/4)x^{-2}$
to $\kappa^2= \kappa_0^2 \approx 38$.  
  The small value of $\kappa_0^2$ means that
driven mode has a significant amplitude at
the distance $r_R$ where the free mode
is excited.

   With both the free perturbation and the quadrupole
driven perturbation present we have
$$
T(r,\phi,t)=
T_0+\Re\big[\delta T_1^f(r)\exp(i\phi-i\omega_1 t)
$$
$$
+~\delta T_1^d(r)\exp(i\phi-i\Omega_* t)\big]~.
\eqno(39)
$$   
The associated flux density 
$
 S \sim T^4$ is
 $$
  T_0^4+4T_0^3\Re\big[\delta T_1^f \exp(i\phi-i\omega_1 t)
+\delta T_1^d\exp(i\phi-i\Omega_* t)\big]
 $$   
 $$
 +6T_0^2 \Re \big\{\delta T_1^f\delta T_1^{d*}
 \exp[-i(\omega_1-\Omega_*)t]\big\} + ...,
 \eqno(40)
 $$
 where the ellipsis denotes terms of the form
 ${\cal O}[\exp(\pm2i\phi)]$ and  ${\cal O}[\exp(\pm 4i\phi)]$ which
 do not cause variations in the total flux.
   It follows from this equation that there are
three  QPO components for a generally  oriented disk.  
Two of the components arise from the above-mentioned
Doppler boost acting first on the term $\delta T_1^f \exp(i\phi -i\omega_1t)$,
which gives a frequency $\omega_1$ component
in the total flux, and second
on the term $\delta T_1^d \exp(i\phi -i\Omega_*t)$, which gives
a frequency $\Omega_*$  (the star's rotation frequency) in
the total flux.
    The third component arises for the final term in
equation (37) and has a frequency $\omega_1 -\Omega_*$.    
The component at $\Omega*$ is however
usually absent (van der Klis 2006) so that we do not
consider this case further.

    For the case of an octupole field component $B^{v2}$
where $m=2$ we find two Lindblad resonances for
$\omega_r/2\pi =\Omega_*/2\pi =300$ Hz and $\omega_i=0$
as shown in Figure 7.   One resonance is
at $r_{Li}/r_S =4.11$ and the other at $r_{Lo}/r_S =4.28$.
  The driven motion at $r_{Li}$ is important here because
this is close to the radius of the unstable free perturbation $r_R$.  
    For this case
$$
T(r,\phi,t)=
T_0+\Re\big[\delta T_1^f(r)\exp(i\phi-i\omega_1 t)
$$
$$
+~\delta T_2^d(r)\exp(2i\phi-i\Omega_* t)\big]~.
\eqno(41)
$$ 
The associated flux density  is
 $$
  T_0^4+4T_0^3\Re\big[\delta T_1^f \exp(i\phi-i\omega_1 t)
+\delta T_2^d\exp(2i\phi-i\Omega_* t)\big]
 $$   
 $$
 +6T_0^2 \Re \big\{\delta T_1^f\delta T_2^{d*}
 \exp[-i\phi-i (\omega_1-\Omega_*)t]\big\} + ...~.
 \eqno(42)
 $$  
 In this case we have just two frequency components:
 The first is at the frequency $\omega_1$ of the
 unstable free perturbation as a result of the Doppler
 boost acting on the term $\delta T_1^f\exp(i\phi-i\omega_1t)$.
 The second is at the frequency $\omega_1 - \Omega_*$
 as a result of the Doppler boost acting on 
 the term  $\delta T_1^f\delta T_2^{d*} 
 \exp[-i\phi-i (\omega_1-\Omega_*)t]$.
   The Doppler boost acting on the remaining
term in equation (39),  $\delta T_2^d\exp(2i\phi-i\Omega_* t)$,
causes no variation in the total flux.

\begin{inlinefigure}
\centerline{\epsfig{file=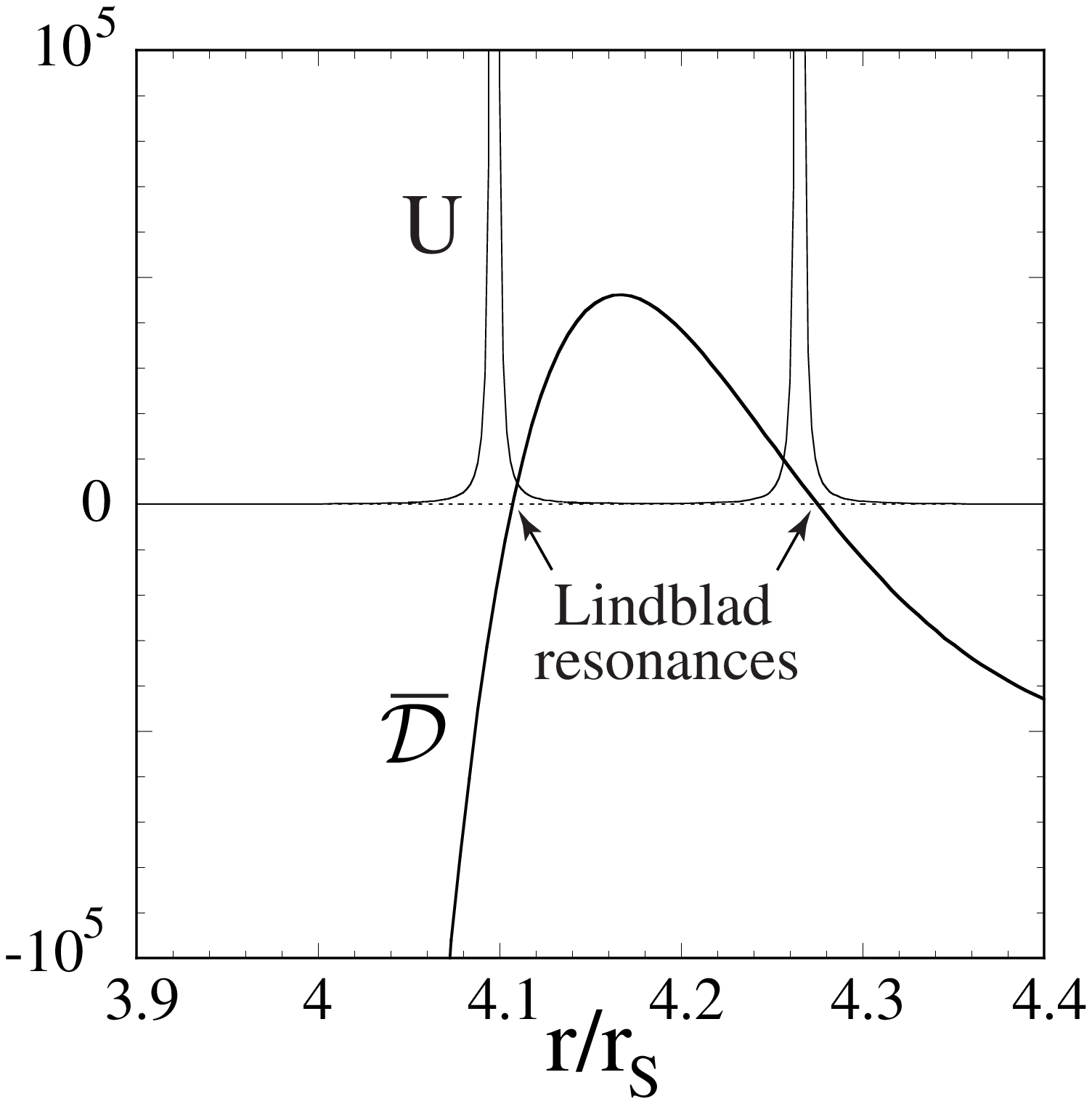,
height=2.9in,width=3.in}}
\epsscale{0.8}
\figcaption{Radial dependence of the 
Lindblad factor $\overline{\cal D}$ (arbitrary scale)  and
the effective potential $U$ (multiplied by $r_S^2$) for
$\omega_r/2\pi=300$ Hz, $\omega_i=0$,
and $m=2$ and other conditions the same as
in Figure 5.
}
\end{inlinefigure}

\subsection{Driven Modulated Perturbations}

     Here we consider the  driven modulated perturbations
of the flow which result from the
octupole component of the star's
magnetic field $B^{v2}$ discussed in \S 2.3.
  Two Lindblad resonances are again found at
radii  similar to the case of Figure 7.   The inner
radius $r_{Li}$ is important here because it
is close to the radius $r_R$ of the unstable
free perturbation.
   For the driven modulated octupole case we have
$$
T(r,\phi,t)=
T_0+\Re\big\{\delta T_1^f(r)\exp(i\phi-i\omega_1 t)
$$
$$
+~\delta T_2^{dm}(r)\exp[2i\phi-i(\Omega_*/2) t]\big\}~.
\eqno(43)
$$ 
The associated flux density  is
 $$
  T_0^4+4T_0^3\Re\big\{\delta T_1^f \exp(i\phi-i\omega_1 t)
+\delta T_2^{dm}\exp[2i\phi-i(\Omega_*/2) t)\big]
 $$   
 $$
 +6T_0^2 \Re \big\{\delta T_1^f\delta T_2^{dm*}
 \exp[-i\phi-i (\omega_1-\Omega_*/2)t]\big\} + ...~.
 \eqno(44)
 $$  
  In this case we again have just two frequency components:
 The first is at the frequency $\omega_1$ of the
 unstable free perturbation as a result of the Doppler
 boost acting on the term $\delta T_1^f\exp(i\phi-i\omega_1t)$.
 The second is at the frequency $\omega_1 - \Omega_*/2$
 as a result of the Doppler boost acting on 
 the term  $\delta T_1^f\delta T_2^{dm*} 
 \exp[-i\phi-i (\omega_1-\Omega_*/2)t]$.
   The Doppler boost acting on the remaining
term in equation (39),  $\delta T_2^{dm}\exp[2i\phi-i(\Omega_*/2)t]$,
causes no variation in the total flux.  
   As mentioned in \S 2.3 the present theory does 
not predict the value of $\delta T_2^{dm}$.   

\section{Nonlinear Effect of Unstable Wave}

      The considered equilibrium disk (\S 3) does not
include accretion.      
     However, the unstable Rossby mode discussed
in \S 4.1 gives rise to accretion in the vicinity of $r_R$,
$$
 \dot{M}_{\rm Ro} =-2\pi r \int_{-h}^h dz~ \Re(\delta \rho \delta  u_r^*)~,
\eqno(45a)
$$
where $h$ is the half-thickness of the disk.
   We evaluate this by taking the most singular terms in   
$\Delta \omega$ in the limit where this quantity almost vanishes
at $r=r_R$.
This gives $\delta \rho =2k_\phi {\cal F} \Psi /(\Delta \omega L_*)$,
and $\delta u_r = -2i k_\phi {\cal F} \Psi /\rho$.    
   Thus,
$$
 \dot{M}_{\rm Ro} =-~{ 8\omega_i h k_\phi^2|{\cal F}|^2 |\Psi|^2
\over \rho |\Delta \omega|^2 L_*}~,
\eqno(45b)   
$$
where $\omega_i>0$ is the growth rate and $L_* <0$.  Thus,
the instability acts increase the mass accretion rate,
$\dot{M}_{\rm Ro} >0$.

    The unstable Rossby mode also gives an inflow of
angular momentum,    
 $$
{F}_{\rm Ro} =-2\pi r \int_{-h}^h dz~ 
\bigg[\Re(\delta \rho \delta  u_r^*)r u_\phi 
+\rho \Re(\delta u_r^* r \delta u_\phi)\bigg]~.
\eqno(46a)
$$
Using the fact that $\rho \delta u_\phi =-k_\phi {\cal F}\Psi p_*^\prime
/(\rho \Omega_\phi \Delta\omega L_*)$, we find
$$
 F_{\rm Ro}=\dot{M}_{\rm Ro}\ell Q~,
\eqno(46b)
$$
where $Q=(1+u_K^2/u_\phi^2)/2$ and $u_K=
r\Omega_K$ is the Keplerian velocity. 
For the case of Figure 6, $ Q=1.884$.
    Because $Q>1$, the Rossby mode transports
inward more specific angular momentum $\ell$ than exists
in the equilibrium.   
  Consequently, a nonlinear effect of the Rossby mode
is to make the positive slope of $\ell(r)$ smaller
in the vicinity of $r_R$.   
       A sufficiently large reduction of $d\ell/dr $ may cause
the  growth rate to decrease.

\section{Conclusions}
 
    We have investigated three  modes of accretion disks 
around rotating magnetized neutron stars which may
explain the separations of the kilo-Hertz
quasi-periodic oscillations (QPO) seen in
low mass X-ray binaries.     This work is a continuation
of the earlier work by LR07
where these modes were identified.
   We develop the theory of compressible, non-barotropic
MHD perturbations of an axisymmetric equilbrium
flow with $\partial/\partial z =0$.
     We assume that there is a maximum
in the angular velocity $\Omega_\phi(r)$ of the accreting material
larger than the angular velocity of the star $\Omega_*$,
and that the fluid is in
approximately circular motion near this
maximum rather than moving
rapidly towards the star or out of the disk plane
into funnel flows.   
    MHD simulations  by Romanova et al. (2002, 2008),
 Long et al. (2005),    
and Kulkarni \& Romanova (2008)    show
this type of flow and $\Omega_\phi(r)$ profile.   
    
   The first mode we find is a Rossby wave instability 
or unstable corotation mode    which is
radially trapped in the vicinity of the maximum of 
a key function $g(r){\cal F}(r)$ at $r_{R}$.  
   We derive a Schr\"odinger type equation
for the perturbation, $\varphi^{\prime \prime} = U(r)\varphi$,
where $U(r)$ is the effective potential.   
    This instability is analogous to that found
earlier by Lovelace \& Hohlfeld (1978), 
Lovelace et al. (1999),    and Li et al. (2000)
and in simulations by Li et al. (2001).
The real part of the angular  frequency of the mode 
is $\omega_r=m\Omega_\phi(r_{R})$, where 
$m=1,2...$ is the azimuthal mode number.
The imaginary part of the frequency $\omega_i$ (the growth rate)
is determined by a Bohr-Sommerfeld quantization 
of the perturbation in the effective potential.   
   We argue that for generally oriented disk,
the Doppler boost of the disk surface emission
from the unstable $m=1$ mode will    give periodic variations
in the total flux with angular frequency $\omega_r$
which we suggest is the higher frequency component
of the twin QPOs  as proposed by LR07
The saturation of the growth of this mode remains to
be analyzed in detail in  future work.

   The second mode,  is a mode driven by 
the rotating, non-axisymmetric quadrupole [$\sim \exp(i\phi)$]
and/or octupole  [$\sim \exp(2i\phi)$] components of the
star's  magnetic field.     It has an angular frequency
equal to the star's angular rotation rate $\Omega_*$.
   This mode is strongly excited near the radius
of the Lindblad resonance which is slightly
outside the radius of the maximum of $g{\cal F}$.   
    When both the first and second modes are present
the nonlinearity of the emission will in general
give a product term with angular frequency $\omega_r-\Omega_*$.  
   For a quadrupole field taking
into account the Doppler boost, we find that the total flux 
has periodic variations at three frequencies, $\omega_r$, $\Omega_*$,
and $\omega_r-\Omega_*$.   
   However, the frequency $\Omega_*$
is usually not observed (van der Klis 2006).
       The situation is  different for an octupole field component again
taking       into account the Doppler boost.
    In this case the total flux 
has periodic variations at two frequencies, $\omega_r$ the upper QPO
and $\omega_r-\Omega_*$ the lower QPO.

    The third mode arises from the interaction of the
flow perturbation with the rotating non-axisymmetric
components  of the star's magnetic field.  
   This interaction arises naturally owing to the
fact that the magnetic force is proportional 
to the gradient of the square of the magnetic field.
   We derive a linear differential equation for this driven modulated
perturbation.  
We find that the angular frequency of the 
perturbation is $\Omega_*/2$ both for the quadrupole and 
the octupole field components.
       In the case of the octupole field component and
taking into account the Doppler boost, we find that the total flux 
has periodic variations at two frequencies, $\omega_r$ the upper QPO
and $\omega_r-\Omega_*/2$ the lower QPO.     
     Because the equation for the driven modulated
perturbations is linear, the amplitude of the motion is
indeterminate.   
   One possibility is that the $\Omega_*/2$ motion
is excited nonlinearly by the rotating non-axisymmetric
field  which has frequency $\Omega_*$.   
  This will be discussed in a future work.   Thus 
present theory does not determine whether the
lower QPO frequency is $\omega_r -\Omega_*$
or $\omega_r - \Omega_*/2$.

\section*{Acknowledgements}

   We thank Mr. Akshay Kulkarni and Prof. Chengmin Zhang for helpful comments.   We thank the Maplesoft Technical Support Team for
valuable assistance.
The authors were supported in part by NASA grant NNX08AH25G and by
NSF grants AST-0607135 and AST-0807129.

\end{document}